\documentclass[review, preprint,12pt]{elsarticle}
\usepackage{amsmath,amssymb,caption,subcaption}
\usepackage{algorithm2e}
\usepackage{upgreek}
\usepackage{color}
\biboptions{numbers,sort&compress}
\usepackage[margin=2cm]{geometry}
\journal{International Journal of Fatigue}
\begin{document}
\begin{frontmatter}
\title{An FFT-based crystal plasticity phase-field model for micromechanical fatigue cracking based on the stored energy density}
\author[add1]{S. Lucarini\corref{corr1}}
\cortext[corr1]{Corresponding authors}
\ead{s.lucarini@imperial.ac.uk}
\author[add2]{F.P.E. Dunne}
\author[add1]{E. Mart{\' i}nez-Pa{\~ n}eda\corref{corr1}}
\ead{e.martinez-paneda@imperial.ac.uk}
\address[add1]{Department of Civil and Environmental Engineering, Imperial College London, London SW7 2AZ, UK}
\address[add2]{Department of Materials, Imperial College London, London SW7 2AZ, UK}
\begin{abstract}
A novel FFT-based phase-field fracture framework for modelling fatigue crack initiation and propagation at the microscale is presented. A damage driving force is defined based on the stored energy and dislocation density, relating phase-field fracture with microstructural fatigue damage. The formulation is numerically implemented using FFT methods to enable modelling of sufficiently large, representative 3D microstructural regions. The early stages of fatigue cracking are simulated, predicting crack paths, growth rates and sensitivity to relevant microstructural features. Crack propagation through crystallographic planes is shown in single crystals, while the analysis of polycrystalline solids reveals transgranular crack initiation and crystallographic crack growth. \\

\end{abstract}


\end{frontmatter}


\section{Introduction}

Fatigue damage in metal alloys is arguably the biggest threat to the service life of engineering components \cite{Suresh1998}. Service life predictions are thus dependent on our ability to model the various stages of fatigue damage, from the nucleation of a fatigue crack to its propagation and unstable failure \cite{Golahmar2023}. In many applications and sectors, such as the aeronautical or automotive industries, the formation of a macroscopic crack corresponds to a very significant portion of the fatigue life. Hence, being able to predict the stages of fatigue crack nucleation and short crack growth is of notable scientific and technological importance. However, these early stages of fatigue damage are very microstructurally-sensitive \cite{Aslan2011}, partially explaining the observed variability in fatigue lives \cite{Schijve2009,Ghonem2010}, and thus require the development of models capable of capturing the microstructural localisation of plastic deformation and how this plastic localisation leads to the formation of microscopic cracks that go on to propagate and become macroscopic cracks. 

Recent modelling approaches for the incubation and early propagation stages of fatigue cracks in polycrystalline metals typically build upon micromechanics-based models and a fatigue driving force definition. For example, the nucleation stage is often described using so-called Fatigue Indicator Parameters (FIPs), which are surrogate measures for fatigue cracking obtained from micromechanical fields \citep{McDowell2010}. FIPs are estimated by using computational polycrystalline homogenisation models, in which the macroscopic response resulting from a specific microstructure can be predicted by solving a boundary value problem on a periodic Representative Volume Element (RVE) of the microstructure. This RVE typically consists of an aggregate of grains modelled using an appropriate crystal plasticity model \citep{Segurado2018}. Although these FIPs-based approaches provide a compelling pathway to account for microstructural effects on fatigue life \citep{Sweeney2013}, they are unable to account for the differentiated regimes of crack initiation, i.e., nucleation and short crack propagation. Understanding these regimes is fundamental for designing fatigue-resistant microstructures and predicting the role of microstructural features such as grain size, distribution, and morphology in fatigue lives \cite{Castelluccio2014}. Recently, efforts have been allocated to enrich computational polycrystalline homogenisation models so as to account for the short fatigue crack growth regime \cite{Rovinelli2017,Long2023}. This is accomplished by establishing suitable fatigue driving force definitions, for example, based on the stored energy density \citep{Xu2021}. However, these analyses have been so far limited to the use of finite element methods, which are limited in their ability to handle large RVEs \cite{Lucarini2019}, and discrete fracture methods such as X-FEM that cannot capture complex cracking conditions (3D, crack branching and coalescence, \textit{etc}.) \citep{Proudhon2016,Wilson2019,Grilli2022}. A step-change in short fatigue crack growth modelling can be achieved by exploiting recent advances in the development of computationally efficient techniques, such as Fast Fourier Transform (FFT)-based methods, and of robust diffuse (phase-field) fracture mechanics models.

The phase-field fracture method \cite{Bourdin2008,Amor2009,PTRSA2021} has gained remarkable popularity in recent years due to its flexibility, robustness, easiness of implementation, as well as its ability to model advanced fracture phenomena such as crack nucleation from arbitrary sites \cite{Lorenzis2022,Navidtehrani2022}, complex crack trajectories \cite{Wu2021,CST2021}, and multiple crack interactions \cite{Borden2012,TAFM2020c}, in arbitrary geometries and dimensions and on the original finite element mesh. Unlike discrete fracture methods, phase-field fracture models smear the interface over a finite domain, using an auxiliary variable to track the evolution of the crack-solid interface without the need for re-meshing or \textit{ad hoc} criteria for crack growth \cite{Bourdin2008,Pons2010}. Although phase-field fracture models were first aimed towards modelling brittle fracture, they were later extended to ductile fracture \cite{Ambati2015,Alessi2018,Aldakheel2021a,Shishvan2021a}, fatigue \cite{Lo2019,Carrara2020,Hasan2021,Golahmar2022} and chemo-mechanical damage \cite{CMAME2018,Boyce2022,Ai2022,Rezaei2023}, among others. Recently, phase-field fracture has also been adopted to model crack initiation and propagation in polycrystalline materials undergoing static loading \cite{Diehl2017,Nguyen2017,Emdadi2021}.

While the combination of phase-field fracture and micromechanical modelling holds great promise, their application to fatigue is limited by the computational cost. An accurate prediction of fatigue behaviour usually requires the simulation of a sufficiently high number of cycles and large RVEs \citep{Boeff2017}, and phase-field simulations require a mesh sufficiently fine to resolve the phase-field length scale. This obstacle can be overcome by using FFT-based methods \citep{Lebensohn2012,Lucarini2022}, which rely on the properties of the derivatives of periodic fields in the Fourier space and exhibit a remarkable numerical performance compared to finite element methods \cite{Lucarini2019}. For a model with $n$ degrees of freedom, FFT-based approaches scale as $n \log n$ while finite element methods (iterative solvers) scale as $\mathcal{O} (n^2)$. Also, FFT solvers suffer from scalability issues or high phase contrast limitations. Recent studies have shown the potential of FFT-based approaches in accelerating phase-field fracture calculations \cite{Chen2019,Ma2020,Xue2022}.

In this work, we combine FFT-based modelling with phase-field fracture to investigate the nucleation and short growth of fatigue cracks in polycrystalline materials. We present a novel formulation whereby a new phase-field fracture driving force is defined based on the crack tip stored energy density concept \cite{Xu2021,Long2023}, extending phase-field fracture models to the simulation of nucleation and short fatigue crack growth. Several case studies are addressed to showcase how the framework is able to predict the nucleation of fatigue cracks, as a result of plastic localisation, and the subsequent expected short crack paths, accounting naturally for microstructural effects and the transition to macroscopic cracking. Additionally, the use of FFT solvers improves the efficiency of the calculations and enables the simulation of sufficiently large 3D RVEs over a relevant number of cycles. The remainder of this manuscript is organised as follows. Section \ref{Sec:Theory} presents the theoretical framework proposed, which combines crystal plasticity, phase-field fracture and fatigue crack tip stored energy density concepts. The numerical procedures employed are described in Section \ref{Sec:Numerics}, including the FFT formulation of the governing equations and the solution algorithm. The results obtained in the 2D and 3D numerical experiments conducted on polycrystalline materials are presented in Section \ref{Sec:Results}, together with the associated discussion. Finally, the manuscript ends with concluding remarks in Section \ref{Sec:Conclusions}.

\section{A crystal plasticity-based phase-field fatigue model}
\label{Sec:Theory}

In this section, the phase-field formulation for short fatigue crack growth is developed within the context of crystal plasticity micromechanics for single- and poly-crystals. The crack phase-field and its gradient are used to regularise the sharp topology of cracks, and crack evolution is driven by the accumulation of elastic and plastic strain energy, as characterised by means of a phenomenological criterion based on the stored energy density parameter. In the following, we describe the constitutive relations for the crystal plasticity model, the phase-field fracture description, and the phenomenological fatigue driving force, together with their interactions.

\subsection{Crystal plasticity constitutive equations}
\label{sec:cp}

The crystal plasticity formulation adopted in this work follows the dislocation-based model developed by Dunne \textit{et al.} \citep{Dunne2007,Dunne2012}. We emphasise that the present micromechanical phase-field fatigue model can be readily used with any other constitutive choice for the crystal plasticity behaviour. For the sake of completeness, we proceed to briefly outline the constitutive laws. Within a continuum setting and considering finite strains and rotations, the total deformation gradient $\boldsymbol{F}$ is assumed to support a multiplicative decomposition into elastic $\boldsymbol{F}_{\mathrm{e}}$ and plastic $\boldsymbol{F}_{\mathrm{p}}$ components that respectively account for lattice distortion and plastic slip, such that
\begin{equation}
\boldsymbol{F} = \boldsymbol{F}_{\mathrm{e}} \boldsymbol{F}_{\mathrm{p}}\text{.}
\end{equation}
Accordingly, the total velocity gradient $\boldsymbol{L}$ can be also decomposed into an elastic part and a plastic part, and can be calculated as:
\begin{equation}
\boldsymbol{L}=\boldsymbol{L}_{\mathrm{e}}+\boldsymbol{F}_{\mathrm{e}} \cdot \boldsymbol{L}_{\mathrm{p}} \cdot \boldsymbol{F}_{\mathrm{e}}^{-1}
\end{equation}
where $\boldsymbol{L}_{\mathrm{e}}$ denotes the elastic velocity gradient and $\boldsymbol{L}_{\mathrm{p}}$ is the plastic velocity gradient. The crystal plasticity description assumes that dislocation slip occurs on slip systems and $\boldsymbol{L}_{\mathrm{p}}$ is calculated as the sum of all the plastic shear strain rate contributions from the active slip systems. Hence, for a total number of slips $\mathrm{N}_{\mathrm{s}}$ (e.g., $\mathrm{N}_{\mathrm{s}}=12$ in FCC lattices), the plastic velocity gradient is given by
\begin{equation}
\boldsymbol{L}_{\mathrm{p}}=\dot{\boldsymbol{F}}_{\mathrm{p}} \boldsymbol{F}_{\mathrm{p}}^{-1}=\sum_{i=1}^{\mathrm{N}_{\mathrm{s}}} \dot{\gamma}^{\mathrm{i}} \mathbf{n}^{\mathrm{i}} \otimes \mathbf{s}^{\mathrm{i}}
\end{equation}
where $\dot{\gamma}^{\mathrm{i}}$ is the shear strain rate for the $\mathrm{i}$-th slip system, and $\mathbf{s}^{\mathrm{i}}$ and $\mathbf{n}^{\mathrm{i}}$ stand for the aligned and normal vectors to the direction of the $\mathrm{i}$-th slip plane. The plastic shear strain rate $\dot{\gamma}^{\mathrm{i}}$ on the $\mathrm{i}$-th slip system can be estimated from the resolved shear stress $\tau^{\mathrm{i}}$ as \cite{Dunne2007}:
\begin{equation}\label{eq:flowrule}
\dot{\gamma}^{\mathrm{i}}=\rho_{\mathrm{m}} b^{2} v_{D} \exp \left(-\frac{\Delta F}{k T}\right) \sinh \left[\frac{\left(\left|\tau^{\mathrm{i}}\right|-\tau_{\mathrm{c}}^{\mathrm{i}}\right) \Delta V}{k T}\right],
\end{equation}
where $\rho_{\mathrm{m}}$ is the density of mobile dislocations, $v_{D}$ is the frequency of attempted dislocation jumps over obstacles, $b$ is the Burgers vector, $\Delta F$ is the thermal activation energy, $k$ is the Boltzmann constant, $T$ is the absolute temperature (295 $K$), $\tau_{\mathrm{c}}^{\mathrm{i}}$ is the critical resolved shear stress (CRSS) on the $\mathrm{i}$-th slip system, and $\Delta V$ is the activation volume. 

The mechanistic slip rule in Eq. \eqref{eq:flowrule} is derived from the notion of dislocation mobility being constrained by the pinning of dislocations, until the thermodynamic driving force causes the dislocation to escape these pinning points, leading to rate-sensitive slip. The CRSS follows a Taylor-based hardening law on these systems, due to the evolving density of statistically stored dislocations (SSDs) and geometrically necessary dislocations (GNDs), such that
\begin{equation}
\tau_{c}^{\mathrm{i}}=\tau_{c 0}^{\mathrm{i}}+\mu b \sqrt{\rho_{SSD}+\rho_{GND}}
\end{equation}
where $\tau_{c 0}^{\mathrm{i}}$ is the initial CRSS, $\mu$ is the elastic shear modulus, and $\rho_{SSD}$ and $\rho_{GND}$ respectively denote the SSD and GND densities. The SSD density is assumed to evolve with accumulated slip and induces hardening on the slip system CRSS. Then, the evolution of the SSD density is defined to vary linearly with the rate of equivalent plastic strain, such that 
\begin{equation}
\dot{\rho}_{SSD}=\lambda \sqrt{\frac{2}{3} \boldsymbol{L}_{\mathrm{p}} : \boldsymbol{L}_{\mathrm{p}}}
\end{equation}
where $\lambda$ is the hardening coefficient. 

GNDs account for permanent lattice curvature, or plastic strain gradients, and can obstruct dislocation glide \cite{Ashby1970,Fleck2015a,JMPS2019}. The Nye tensor $\boldsymbol{\Lambda}$ \citep{Nye1953}, and its characterisation of plastic incompatibility, are used to estimate the GND density. Accordingly, the definition of the Nye tensor $\boldsymbol{\Lambda}$ is given by,
\begin{equation}\label{eq:nye1}
\boldsymbol{\Lambda}=\nabla_0 \times \boldsymbol{F}_{\mathrm{p}}=\sum_{\mathrm{i}=1}^{\mathrm{N}_{\mathrm{s}}} \rho_{Gs}^{\mathrm{i}} \mathbf{b}^{\mathrm{i}} \otimes \mathbf{s}^{\mathrm{i}}+ \rho_{Ge}^{\mathrm{i}} \mathbf{b}^{\mathrm{i}} \otimes \mathbf{t}^{\mathrm{i}}\text{,}
\end{equation}
where $\nabla_0 \times$ denotes the curl in the reference configuration,  $\rho_{\mathrm{G}_{\mathrm{s}}}^{\mathrm{i}}$ and $\rho_{\mathrm{G}_{\mathrm{e}}}^{\mathrm{i}}$ respectively stand for the pure screw and pure edge independent dislocation components, and $\mathbf{b}^{\mathrm{i}}$ and $ \mathbf{t}^{\mathrm{i}}$ are the Burgers vector and the tangent vector to the direction of the $\mathrm{i}$-th slip plane. In the FFC lattice, the number of pure edge and screw dislocation segments equals 12 and 6, respectively. Then, the inherent non-uniqueness of the GND solution is addressed by using the L$_2$-norm scheme, which minimises the sum of the squares of the resulting dislocation densities vectors to obtain the screw and edge components of the dislocation densities, leading to the following scalar GND density value \cite{Cheng2015}:
\begin{equation}
\rho_{GND}=\sqrt{\sum_{\mathrm{i}=1}^{\mathrm{N}_{\mathrm{s}}}\left(\rho_{Gs}^{\mathrm{i}}\right)^{2}+\left(\rho_{Ge}^{\mathrm{i}}\right)^{2}}\text{.}
\end{equation}
Note that the consideration of a smooth plastic deformation gradient field is only valid for small values of plastic deformation and low misorientation boundaries, and implies an artificial increased GND effect on high-angle grain boundaries and on already damage zones.

\subsection{Phase-field modelling for fracture}

The nucleation and growth of cracks are characterised by a phase-field description, whereby an auxiliary field variable (the phase-field $\phi$) is introduced to describe the crack-intact material interface. This field variable $\phi$ exhibits a continuous, smooth transition within a given interval, e.g. $\phi \in \left[ 0, 1 \right]$, taking the lower limit value ($\phi=0$) in the undamaged phase and the upper limit magnitude ($\phi=1$) in fully cracked material points. The phase-field variable is typically assumed to evolve based on the balance between stored and fracture energies, following the energy balance first postulated by Griffith for brittle solids \cite{Griffith1920}. Accordingly, the potential energy of the solid $\Psi$ can be written as the additive decomposition of the mechanical stored energy and the fracture energy, which in the reference configuration reads
\begin{equation}\label{eq:EnergyBalance}
\Psi\left(\boldsymbol{F}, \Gamma_0\right)=\int_{\Omega_0} W(\boldsymbol{F}) \ \mathrm{d} \Omega_0+\int_{\Gamma_0} \mathcal{G}_c^0 \ \mathrm{d} \Gamma_0
\end{equation}
where $W$ denotes the stored strain energy per unit volume and $\mathcal{G}_c^0$ is the critical fracture energy per unit area. The phase-field regularised form of Eq. (\ref{eq:EnergyBalance}) introduces a crack density function $\Gamma_c$ to provide a volumetric approximation of the energy contribution from the crack surface and thus enable predicting the evolution of cracks based on an exchange between the stored and fracture energies \cite{Bourdin2008,Kuhn2010}; 
\begin{equation}
\tilde{\Psi}\left(\boldsymbol{F}, \phi\right)=\int_{\Omega_0} g\left(\phi\right) W\left(\boldsymbol{F}\right) \ \mathrm{d} \Omega_0+\int_{\Omega_0} \mathcal{G}_c^0 \Gamma_c\left(\phi, \nabla \phi\right) \ \mathrm{d} \Omega_0
\end{equation}
Here, $g(\phi)$ is the degradation function that reduces the material stiffness, which we assume to be of a quadratic form:
\begin{equation}\label{eq:deg}
g\left( \phi \right)= \left( 1 - \phi \right)^2 + k
\end{equation}
with $k$ being a residual numerical stiffness that prevents the mechanical equilibrium system of equations to become singular. In this work, a value of $k= 10^{-5}$ is assumed and this numerical term is henceforth dropped for simplicity. The crack density function is often assumed to be given by \cite{Bourdin2008}
\begin{equation}
\Gamma_c\left(\phi, \nabla \phi\right)=\frac{1}{2 \ell}\left( \phi^2 + \ell^2 \left|\nabla \phi \right|^2 \right)
\end{equation}
where $\ell \in \mathbb{R}^{+}$ is a length scale parameter that determines the width of the regularised crack.

In extending this theory to ductile materials, an effective plastic work contribution $W_p$ is incorporated into the total potential energy of the solid \citep{Miehe2015}:
\begin{equation}\label{eq:tot}
\tilde{\Psi}\left(\boldsymbol{F}, \phi, \boldsymbol{\upalpha} \right) = \int_{\Omega_0} g(\phi) \left[ W_\mathrm{e}\left(\boldsymbol{F}_{\mathrm{e}}\right) + W_\mathrm{p}(\boldsymbol{F},\boldsymbol{\upalpha}) \right]  + \mathcal{G}_c^0 \Gamma_c\left(\phi, \nabla \phi\right) \ \mathrm{d} \Omega_0
\end{equation}
where $\boldsymbol{\upalpha}$ is the vector of internal variables and $W_\mathrm{e}$ is the elastic stored energy per unit volume. Note that the plastic energy contributes to crack growth and is degraded in an analogous manner to its elastic counterpart. The role of plastic strains in contributing to damage evolution can be readily observed by deriving the Euler-Lagrange equation from Eq. \eqref{eq:tot}, rendering the following strong form:
\begin{equation}\label{eq:strpf}
\left(\phi-\ell^{2} \nabla^{2} \phi\right) - 2(1-\phi)\mathcal{S} \left( \text{state}\left(\boldsymbol{F}, \boldsymbol{\upalpha}\right)\right)=0
\end{equation}
where $\mathcal{S} \left( \text{state}\left(\boldsymbol{F}, \boldsymbol{\upalpha}\right)\right)$ is a crack driving state function, which for the conventional phase-field fracture model equals $\mathcal{S} \left( \text{state}\left(\boldsymbol{F}, \boldsymbol{\upalpha}\right)\right)=\ell/\mathcal{G}_c^0\left(W_{\mathrm{e}}+W_{\mathrm{p}}\right)$. This fracture driving force is here reformulated to extend phase-field fracture modelling to the analysis of microstructural fatigue crack nucleation and growth.

\subsection{A fracture driving force for microstructural fatigue cracks}

Our choice for the phase-field fracture driving force relies on the stored energy density concept, as its ability to predict microstructural fatigue crack nucleation and growth has been demonstrated \cite{Wilson2019,Chen2018}. This approach is phenomenological yet mechanistic, relating the plastic work and the dislocation density with the current fatigue damage state of the material. In the crystal plasticity formulation, the stored energy density is determined from the fraction of the plastic work that is stored in the local dislocation structure and normalised by the length scale over which the energy is stored (the dislocation mean free path), rendering
\begin{equation}
G_s=\int \frac{\xi \boldsymbol{\sigma} : \mathrm{d} \boldsymbol{\varepsilon}^{\mathrm{p}}}{\sqrt{\rho_{SSD}+\rho_{GND}}}
\end{equation}
Fatigue cracking will then occur when the stored energy density reaches a critical value, $G_s \rightarrow G_{crit}$. We incorporate this concept in the context of phase-field fracture by defining a new crack driving state function  $\mathcal{S} \left( \text{state}\left(\boldsymbol{F}, \boldsymbol{\upalpha}\right)\right)$ and a new stored energy density-based threshold $W_{\mathrm{crit}}$.

The condition $G_s \rightarrow G_{crit}$ is accounted for by defining a threshold condition for the onset of damage initiation, which is then incorporated into the phase-field balance equation. Thus, a variable $W_{\mathrm{crit}}$ is defined as the sum of elastic and plastic energies at the time when the stored energy density reaches the critical value; i.e.,
\begin{equation}\label{eq:crit}
W_{\mathrm{crit}}=
\left. \left( 
\max_{\tau \in \left[ 0,t \right] } 
\left( W_{\mathrm{e}}^{+} \right) + W_{\mathrm{p}} 
\right) \right|_{G_s =G_{crit}} ,
\end{equation}
where $\max_{\tau \in \left[ 0,t \right] } \left( W_{\mathrm{e}}^{+} \right)$ stands for the maximum value reached by the tensile part of the elastic strain energy, where a spectral decomposition of the elastic tensor is applied \cite{Miehe2016}. That is, 
\begin{equation}
    W_{\mathrm{e}}^{+} (\boldsymbol{\varepsilon})=\boldsymbol{\varepsilon}_{+}^e : \mathbb{C} : \boldsymbol{\varepsilon}_{+}^e  \,, \hspace{1mm}\text{with}\hspace{2mm}  \boldsymbol{\varepsilon}_{+}^e=\sum_{i=1}^{3}\left\langle\varepsilon_{i}^e\right\rangle_{+} \mathbf{n}_{i} \otimes \mathbf{n}_{i}
    \label{eq:decMiehe}
\end{equation}
where $\left\langle\varepsilon_{i}^e\right\rangle_{+}$ and $\mathbf{n}_{i}$ are the positive eigenvalues and eigenvectors of the Eulerian logarithmic elastic strain tensor (i.e., $\boldsymbol{\varepsilon}^e = 1/2 \ln \left( \boldsymbol{F}_{\mathrm{e}} \boldsymbol{F}_{\mathrm{e}}^T \right)$), and $\mathbb{C} $ denotes the anisotropic fourth order elastic stiffness tensor of the crystal lattice, which contains cubic symmetries.

Then, a new crack driving state function is defined as 
\begin{equation}\label{eq:zeta}
\mathcal{S}\left(\text{state}\left(\boldsymbol{F}, \boldsymbol{F}_{\mathrm{p}}, \boldsymbol{\upalpha}\right)\right)=\frac{\ell W_{\mathrm{crit}}}{\mathcal{G}_c^0}\left\langle\frac{W_{\mathrm{e}}^{+}+W_{\mathrm{p}}}{W_{\mathrm{crit}}}-1\right\rangle,
\end{equation}
where $\langle x\rangle:=(x+|x|) / 2$ are the Macaulay brackets. Inserting Eq. (\ref{eq:zeta}) in the phase-field balance equation \eqref{eq:strpf} delivers a model capable of nucleating and growing cracks based on the mechanisms driving microstructural fatigue damage. Two particularly relevant features can be noted. First, while other phase-field fracture models have exploited the concept of a damage threshold, it has always been assumed to be a constant value. Here, the threshold $W_{crit}$ is highly dependent on the local stored energy density. Second, a dimensionless parameter $\zeta=\ell W_{\mathrm{crit}}/\mathcal{G}_c^0$ can be identified in Eq. \eqref{eq:zeta}, which triggers the activation of damage due to the elasto-plastic contribution and governs the post-nucleation behaviour. The phase-field damage tends to increase sharply for high values of $\zeta$, whereas low propagation rates are obtained for a small $\zeta$. Moreover, $\zeta$ is dependent on the local stored energy density parameter, suggesting the need for experimental calibration to define the most accurate $W_{\mathrm{crit}}/\mathcal{G}_c^0$ relation. In this work, $\mathcal{G}_c^0$ is assumed to be proportional to $W_{crit}$ and, for simplicity, $\zeta$ is assumed to be constant and equal to 1 (i.e., the magnitude of $\mathcal{G}_c^0$ can be estimated for the choices of $\ell$ and $W_{\mathrm{crit}}$).

\section{Numerical procedures}
\label{Sec:Numerics}

This section describes the details of the numerical implementation. The FFT-based solver is employed to solve both the mechanical boundary value problem and the phase-field governing equation predicting the onset and growth of fatigue cracks. An alternate minimisation method is used to calculate the solution of the resulting coupled non-linear system of equations. The implementation is carried out in the \texttt{FFTMAD} package \citep{Lucarini2019}, which is wrapped with Abaqus \texttt{UMAT} subroutines for the calculation of the constitutive equations.

\subsection{FFT-based formulation of the governing equations}

For both the deformation and phase-field equations, we address the periodic boundary value problem using the FFT spectral method and assuming quasi-static conditions. The field equations are posed in a periodic parallelepiped domain $\Omega$ with dimensions $L_1\cdot L_2\cdot L_3$, accounting for both periodic geometry and periodic boundary conditions. 

We start by formulating the strong form of the coupled problem. For the linear momentum balance in the reference configuration, considering the degradation function (\ref{eq:deg}) and a hybrid solution scheme \cite{Ambati2015a}, one gets
\begin{equation}\label{eq:str1}
    \nabla_0 \cdot \left[ \left( 1- \phi \right)^2 \boldsymbol{P} \left( \boldsymbol{F}, \boldsymbol{F}_{\mathrm{p}}, \boldsymbol{\upalpha} \right) \right] = 0
\end{equation}
where $\boldsymbol{P}$ is the first Piola-Kirchhoff stress tensor and $\nabla_0\cdot$ stands for the divergence operator in the reference configuration. Equation \eqref{eq:str1} is solved under the constraint of an evolution law for the internal variables $\boldsymbol{\upalpha}$, such that
\begin{equation}
\dot{\boldsymbol{\upalpha}}=\dot{\boldsymbol{\upalpha}} \left( \boldsymbol{F}, \boldsymbol{F}_{\mathrm{p}}, \boldsymbol{\upalpha}\right)
\end{equation}
which follows the crystal plasticity constitutive laws (Section \ref{sec:cp}). On the other side, the phase-field balance reads
\begin{equation}\label{eq:str2}
    \left( \phi - \ell^2 \nabla_0^2\phi \right) - 2 \left( 1- \phi \right) \frac{\ell W_{\mathrm{crit}}}{\mathcal{G}_c^0}\left\langle\frac{W_{\mathrm{e}}^{+}+W_{\mathrm{p}}}{W_{\mathrm{crit}}}-1\right\rangle = 0
\end{equation}
where $\nabla_0^2$ is the Laplacian operator in the reference configuration and the newly defined crack driving state function $\mathcal{S}$ (\ref{eq:zeta}) has been introduced. 

Within the Fourier method, the domain $\Omega$ is discretised in a voxelised regular grid containing $N_1\cdot N_2\cdot N_3$ voxels. The fields involved in the problem will then be represented by their value at the centre of each voxel. The Fourier space is discretised in the same number of frequencies, and the Fourier transform of a function defined in  $\Omega$ is obtained by the Discrete Fourier Transform of the discrete field and computed using the FFT algorithm. The Fourier space discretisation is defined by the frequency vectors
\begin{equation}
\boldsymbol{\upxi}=\xi_i=2\pi \frac{n_i- \lfloor N_i /2 \rfloor}{L_i}\quad \text{ for } \quad n_i =0, \dots, N_i -1 
\end{equation}
where $n_i$ is the frequency number and $L_i$ and $N_i$ are, respectively, the length of the domain $\Omega$ in the $i$-th direction and the number of voxels in which it is discretised.

Equations \eqref{eq:str1} and \eqref{eq:str2} are then translated to the Fourier space by using frequencies for the spatial derivatives, rendering
\begin{equation}\label{eq:momentum}
    \mathcal{F} \left\{ \left( 1-\phi\right)^2 \boldsymbol{P} \left( \boldsymbol{F}, \boldsymbol{F}_{\mathrm{p}}, \boldsymbol{\upalpha}\right) \right\} \cdot i\boldsymbol{\upxi} = 0
\end{equation}
\begin{equation}\label{eq:phase}
    \left(\phi - \ell^2\mathcal{F}^{-1} \left\{ i^2\boldsymbol{\upxi}^2 \mathcal{F} \left\{ \phi \right\} \right\}  \right) - 2 \left( 1- \phi \right) \frac{\ell W_{\mathrm{crit}}}{\mathcal{G}_c^0}\left\langle\frac{W_{\mathrm{e}}^{+}+W_{\mathrm{p}}}{W_{\mathrm{crit}}}-1\right\rangle = 0
\end{equation}
where $\mathcal{F}^{-1}$ and $\mathcal{F}$ stand for the inverse and direct Fourier transform of a real-valued function and $i$ is the imaginary unit. 
 
For the linear momentum balance, a Newton-Raphson procedure is performed where the linearised form of Eq. \eqref{eq:momentum} is posed. For a given iteration, the deformation gradient is defined as a function of the displacement field in the Fourier space $\widehat{\mathbf{u}}_i$ and the macroscopic deformation gradient $\overline{\boldsymbol{F}}$ by 
\begin{equation}\label{eq:fdef}
\boldsymbol{F}_i= \mathcal{F}^{-1} \left\{  \widehat{\mathbf{u}}_i \otimes \boldsymbol{\upxi} \right\} + \overline{\boldsymbol{F}}
\end{equation}
Then applying corrections, $\delta \widehat{\mathbf{u}}$, to the displacement field following $\widehat{\mathbf{u}}=\widehat{\mathbf{u}}_i + \delta \widehat{\mathbf{u}}$, the linearised mechanical equilibrium reads
\begin{equation}\label{eq:newton}
    \mathcal{F} \left\{ \left( 1- \phi \right)^2 \left. \frac{\partial \boldsymbol{P}}{\partial \boldsymbol{F}}\right|_i :  \mathcal{F}^{-1} \left\{  \delta \widehat{\mathbf{u}} \otimes i\boldsymbol{\upxi} \right\} \right\} \cdot i\boldsymbol{\upxi}  = 
    -\mathcal{F} \left\{ \left( 1- \phi \right)^2 \boldsymbol{P}_i \right\} \cdot i\boldsymbol{\upxi}
\end{equation}
being $\boldsymbol{P}_i=\boldsymbol{P}\left( \boldsymbol{F}_i,\boldsymbol{\upalpha}_i\right)$, and requiring the definition of a consistent tangent $\mathbb{K}=\left. \frac{\partial \boldsymbol{P}}{\partial \boldsymbol{F}}\right|_i$. For strain-controlled tests, $\overline{\boldsymbol{F}}$ is the imposed macroscopic strain, acting as the input boundary condition of the problem. This linear system of equations is reduced by accounting for the real Fourier transform symmetries and solved by means of a Krylov solver (GMRES with a relative tolerance of $tol_{lin}=10^{-5}$), similar to the DBFFT method proposed in Ref. \cite{Lucarini2019b}, where a stress/mixed controlled extension is derived. The Newton-Raphson procedure is considered converged when the gradient in the reference configuration of the displacement correction is sufficiently small $\lVert\nabla_0\delta \mathbf{u}\rVert_{\infty}/\lVert\boldsymbol{F}_i-\boldsymbol{F}_t\rVert_{\infty}< 5 \cdot 10^{-3}$. This scheme requires preconditioning to solve the system of equations given by \eqref{eq:phase} and \eqref{eq:newton}. The preconditioners used in this case are operators defined in the Fourier space, as follows. For the linearised mechanical problem, the preconditioner operator $\widehat{\mathbb{M}}_\mathbf{u}$ reads as
\begin{equation}\label{eq:precf}
    \widehat{\mathbb{M}}_\mathbf{u} \left( \boldsymbol{\upxi} \right) (\ast)=
    \left[ \boldsymbol{\upxi} \cdot \overline{\mathbb{K}} \cdot \boldsymbol{\upxi} \right]^{-1} \cdot \ast 
\end{equation}
where $\overline{\mathbb{K}}$ is the volume averaged consistent tangent tensor
\begin{equation}\label{eq:avstiff}
    \overline{\mathbb{K}}=\frac{1}{V_\Omega} \int_{\Omega} \mathbb{K}\left(\mathbf{x}\right) \mathrm{d}\Omega
\end{equation}
and $V_\Omega$ represents the volume of the entire domain $\Omega$. For the phase-field balance, the preconditioner operator $\widehat{\mathbb{M}}_\phi$ is given by
\begin{equation}\label{eq:precph}
    \widehat{\mathbb{M}}_\phi \left( \boldsymbol{\upxi} \right) (\ast)=
    \left[ 1 + \boldsymbol{\upxi}^2 \right]^{-1} \ast 
\end{equation}
and is applied to Eq. \eqref{eq:phase}.

Finally, the Nye tensor $\boldsymbol{\Lambda}$ (Eq. \ref{eq:nye1}) is computed using the Fourier properties of derivation, so that its expression becomes simpler in the Fourier space, i.e.
\begin{equation}\label{eq:nyeindexfourier}
    \boldsymbol{\Lambda}=\mathcal{F}^{-1} \left\{ i\boldsymbol{\xi} \times \mathcal{F} \left\{\boldsymbol{F}_\mathrm{p}^T\right\} \right\}
    \text{,}
\end{equation}
where $\times$ denotes the cross product. The Nye tensor $\boldsymbol{\Lambda}$ is generally computed inside each crystal domain and then the dislocation density is obtained in the polycrystal homogenisation process. This approach can be problematic due to the singularity of the plastic strain fields over the crystal boundaries. In this study, the evaluation of the plastic strain incompatibility is computed over the entire domain, assuming that the plastic deformation gradient field is smooth, as in Ref. \citep{Haouala2020}. In addition, a regularisation technique is applied for the calculation of Nye's tensor to avoid discretisation effects. Thus, following Ref. \cite{Magri2022}, we estimate the spatial variation of the plastic deformation gradient by solving the following equation in the Fourier space,
\begin{equation}
    \widetilde{\boldsymbol{F}_\mathrm{p}}- \ell_{\boldsymbol{F}_{\mathrm{p}}} \nabla^2_0 \widetilde{\boldsymbol{F}_\mathrm{p}} = \boldsymbol{F}_\mathrm{p}
\end{equation}
where $\ell_{\boldsymbol{F}_\mathrm{p}}$ is the relevant characteristic length of the regularisation.

Note that the proposed FFT framework might exhibit Gibb's oscillation phenomena due to the use of a standard FFT derivative and the inclusion of high phase contrast in the domain. Mitigation mechanisms, such as the use of discrete derivative definitions \citep{Willot2015} or FFT preconditioned FEM schemes \citep{Ladecky2023}, could be employed.

\subsection{Non-linear alternating minimisation}

We solve the non-linear coupled system of equations by using a non-linear alternating minimisation technique (algorithmic scheme described in Algorithm \ref{alg:alternate}), which encompasses an implicit time discretisation approach. Thus, for a given time increment, Eqs. \eqref{eq:newton} and \eqref{eq:phase} are solved in a staggered manner and the process is repeated until reaching convergence for a tolerance of $tol=5 \cdot 10^{-3}$ based on the maximum relative correction of the solution of the solutions (displacements and phase-field) at each iteration (see Alg. \ref{alg:alternate}). Additionally, upon the resolution of Eq. \eqref{eq:phase}, an extra constraint is imposed within the solver to enforce damage irreversibility by assuming $\dot{\phi} \ge 0$, which is imposed explicitly enforced in the solution field within each global iteration.

\SetKwComment{Comment}{/* }{ */}
\RestyleAlgo{ruled}
\begin{algorithm}[H]
\caption{Non-linear alternating minimisation of a time increment}\label{alg:alternate}
\SetAlgoLined
\KwData{$\boldsymbol{F}_t$, $\overline{\Delta\boldsymbol{F}}$, $\Delta t$, $tol$, $\phi_t$}
\KwResult{$\boldsymbol{F}_{t+\Delta t}$, $\phi_{t+\Delta t}$}
$\boldsymbol{F}_0 \gets \boldsymbol{F}_t + \overline{\Delta\boldsymbol{F}}$\\
$\phi^0 \gets \phi_t$\\
$i \gets 1$\\
\While{$\frac{\lVert\boldsymbol{F}_i-\boldsymbol{F}_{i-1}\rVert_{\infty}}{\lVert\boldsymbol{F}_i-\boldsymbol{F}_t\rVert_{\infty}}>tol$ \textbf{\normalfont \textbf{ or } } $\frac{\lVert\phi_i-\phi_{i-1}\rVert_{\infty}}{\lVert\phi_i-\phi_t\rVert_{\infty}}>tol$}{
$\delta\mathbf{u} \gets $ Solve Eq. \eqref{eq:momentum}  by Newton-Krylov solver with a tolerance of $tol$ for $\boldsymbol{F}=\boldsymbol{F}_{i-1}$ and $\phi=\phi_{i-1}$\\
$\boldsymbol{F}_i \gets  \boldsymbol{F}_{i-1}+\nabla_0\delta\mathbf{u}$\\
$\phi_i \gets $Solve Eq. \eqref{eq:phase}  by Krylov solver with a tolerance of $tol$ for $\boldsymbol{F}=\boldsymbol{F}_{i}$\\
\textbf{\normalfont \textbf{if } } $\phi_i<\phi_t$ \textbf{\normalfont \textbf{ then } } $\phi_i=\phi_t$ \textbf{\normalfont \textbf{ endif } }\Comment*[r]{Pointwise}
$i \gets i+1$\\}
$\boldsymbol{F}_{t+\Delta t} \gets \boldsymbol{F}_i$\\
$\phi_{t+\Delta t} \gets \phi_i$\\
\end{algorithm}

\section{Numerical examples and discussion}
\label{Sec:Results}

In this section, we present and discuss the results obtained with the proposed framework in a variety of numerical cyclic loading experiments, including single crystal, bicrystal and polycrystalline specimens. In all cases, the computational domains are discretised in regular grids of voxels where the characteristic voxel length is chosen to be $\ell_{voxel}=0.78$ $\mu$m. This is in line with previous works \citep{Xue2022} and enables establishing a good resolution of the geometry and morphology of the grain structures ($\approx$1000 voxels/grain). Without loss of generality, we introduce the mechanical load by considering strain-controlled uniaxial cyclic testing conditions with a strain amplitude of $\varepsilon_{max}=2$\%, strain ratio $R_{\varepsilon}=0$, and strain rate $\dot{\varepsilon}=1\cdot 10^{-3}$ s$^{-1}$. Note that these loading conditions lead to a more homogeneous distribution of plasticity relative to high-cycle fatigue, where localization effects are prominent, and large RVEs containing microstructural defects are necessary for accurate analyses. However, the framework could be readily used in other loading regimes. Numerical experiments are conducted on FCC crystal structures, with the crystal plasticity parameters being the ones corresponding to a CMSX-4 single crystal at room temperature (see Ref. \cite{Chen2017}); these are listed in Table \ref{tab:cp}. The calculation of Nye's tensor is carried out using a characteristic regularisation length of 2 voxels ($\ell_{\boldsymbol{F}_\mathrm{p}}=1.56$ $\upmu$m), which has shown to be the minimum required to vanish discretisation effects \cite{Magri2022}. The phase-field parameters are chosen so as to ensure that a sufficient number of loading cycles take place, so that stabilised fatigue loops are reached before crack formation and propagation. Fatigue crack nucleation phenomena can be predicted with a few simulation cycles and the conclusions extracted from a stabilised fatigue response can be applied to total life estimates \cite{Stinville2022}. Specifically, we consider $G_{crit}=4$ Jm$^{-2}$ and take the phase-field length scale to be $\ell=2.34$ $\upmu$m, three times larger than the characteristic voxel length so as to ensure discretisation-independent results \cite{Chen2019}. Note that for quantitative estimates of fatigue life, extrapolation schemes are needed.

\begin{table}[ht!]\centering
\caption{Parameters used for the crystal plasticity model.}\label{tab:cp}
\begin{tabular}{lr}
\hline
Parameter & Magnitude \\
\hline
Elastic constant $C_{11}$ [GPa] & 250 \\
Elastic constant $C_{12}$ [GPa] & 161 \\
Elastic constant $C_{44}$ [GPa] & 129 \\
Initial CRSS $\tau_{c0}$ [MPa] & 350 \\
Burgers vector $b$ [mm] & 3.5$\times$10$^{-7}$ \\
Frequency of attempted dislocation jumps $v_D$ [s$^{-1}$] & 1$\times$10$^{11}$ \\
Boltzmann constant $k$ [JK$^{-1}$] & 1.4$\times$10$^{-23}$ \\
Thermal activation energy $\Delta F$ [J] & 4.9$\times$10$^{-20}$ \\
Density of mobile dislocations $\rho_{m}$ [mm$^{-2}$] & 5.0$\times$10$^{6}$ \\
Hardening coefficient $\lambda$ [mm$^{-2}$] & 150$\times$10$^{6}$ \\
\hline
\end{tabular}
\end{table}

\subsection{Failure of a single crystal plate as a function of the lattice orientation}

\begin{figure}[ht!]\centering
\includegraphics[width=0.8\textwidth]{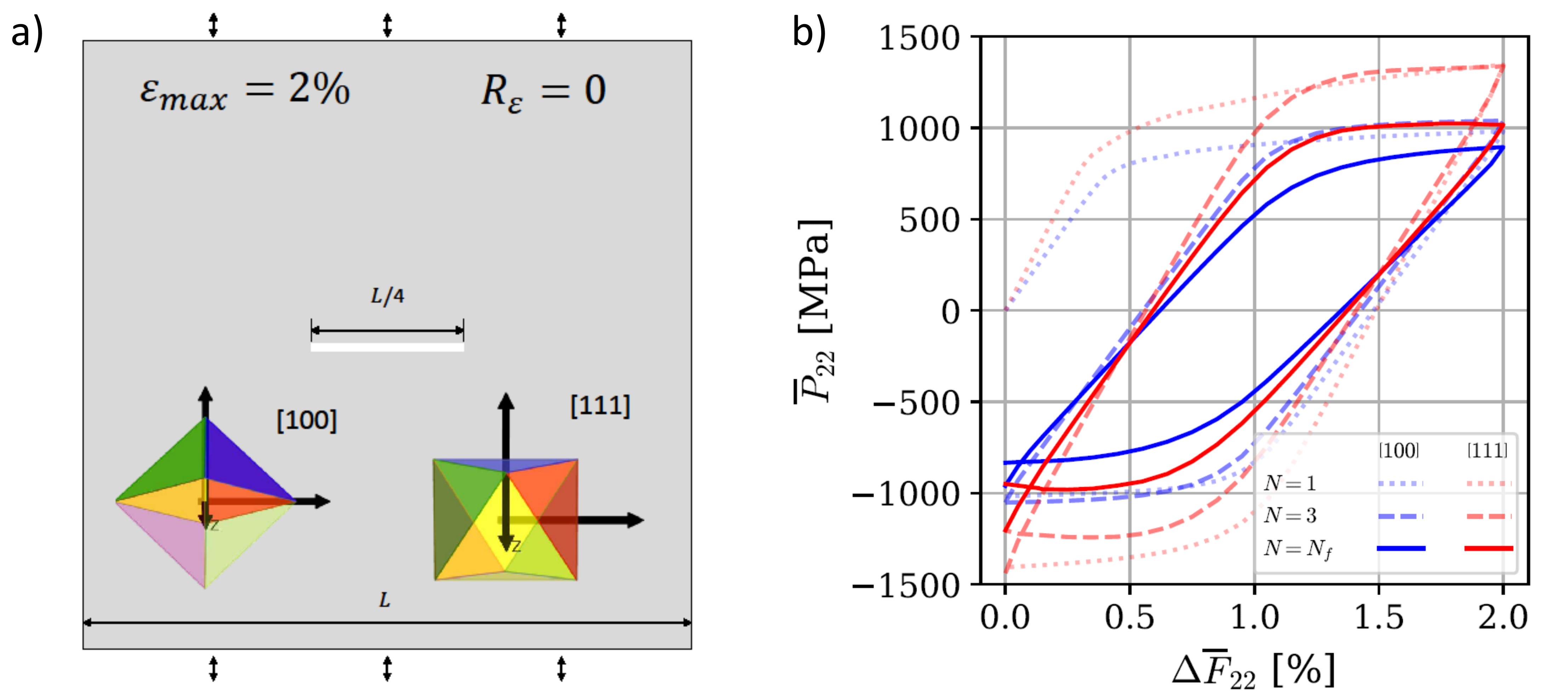}
\caption{Fatigue failure of a single crystal plate as a function of the orientation: (a) geometry of the sample, boundary conditions and sketch of the crystal orientations, and (b) macroscopic stress-strain results (2-vertical direction) obtained for both orientations ([100] and [111]) at the first cycle, the third cycle and the failure cycle.}
\label{fig:sxgeo}
\end{figure}

The first case study involves the initiation of growth and subsequent propagation of microstructural fatigue cracks in a single crystal plate as a function of the crystal orientation. The sample, depicted in Fig. \ref{fig:sxgeo}a, has dimensions $L_x=200$ $\upmu$m $\times$ $L_y=200$ $\upmu$m $\times$ $L_z=0.78$ $\upmu$m and is discretised using $N_x=256$ $\times$ $N_y=256$ $\times$ $N_z=1$ voxels. An initial crack of length 50 $\upmu$m is introduced in the centre of the sample, perpendicular to the loading direction (mode I fracture). This through-thickness crack was introduced geometrically for numerical purposes as it requires handling high levels of plasticity in the damaged region from the begging of the analysis if pre-damage is introduced. Specifically, the geometrical crack is introduced by defining a region of linear elastic material with a stiffness $10^{-5}$ times the original stiffness of the crystal. This region spans the voxels located in the domain $\left( 3L_x/8,5L_x/8 \right)$, $\left( L_y/2-L_y/N_y,L_y/2 + L_y/N_y \right)$. Plane stress conditions are assumed. Two different orientations are analysed: (i) a single crystal oriented in the [100] direction, so that there is a preferential slip system at 45 degrees, and (ii) a rotated configuration, so that the [111] direction is aligned with the load and perpendicular to the crack. The crystal orientations are shown in Fig. \ref{fig:sxgeo}, together with a sketch of the sample, the loading conditions, and the macroscopic stress-strain loops obtained for both orientations.

\begin{figure}[ht!]\centering
\includegraphics[width=0.8\textwidth]{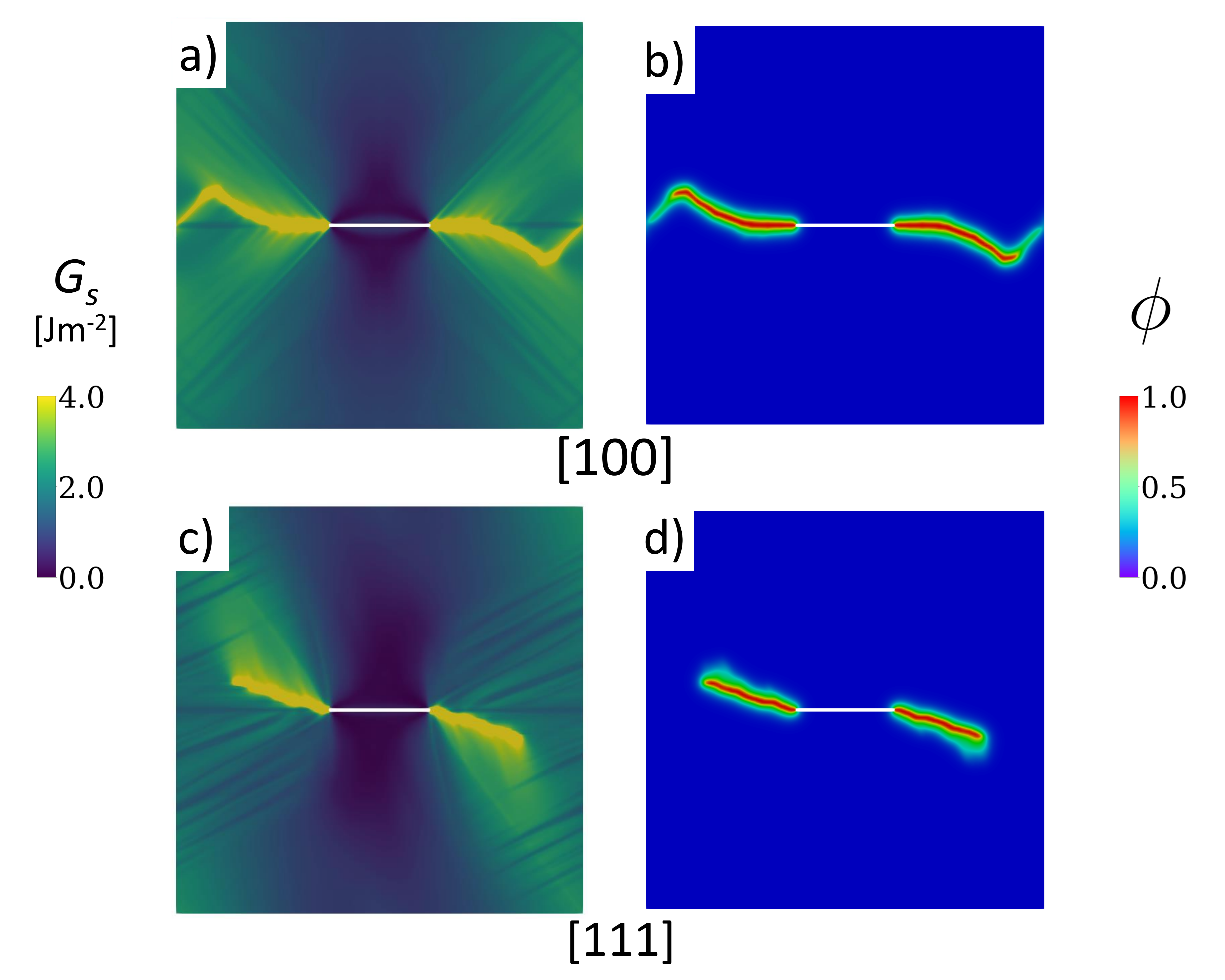}
\caption{Fatigue failure of a single crystal plate as a function of the orientation: [100] (top) versus [111] (bottom). Subfigures (a) and (c) show the stored energy density $G_s$ contours, while subfigures (b) and (d) provide the predicted crack trajectories, as described by the phase-field contours. The results are shown at the end of the failure cycle $N_f$.}
\label{fig:sx}
\end{figure}

The stress-strain behaviour is shown until the failure cycle $N_f$, with $N_f=6$ for the [100] case and $N_f=4$ for the [111] analysis. The [111] sample (red lines) exhibits a stiffer and stronger response across the entire cycle history. In both cases, the crystal cyclic behaviour quickly stabilises, with the $N=3$ results corresponding to the stabilised stress-strain curves. Some influence of damage is observed in the failure cycle before the crack propagates throughout the specimen. It is important to note that the observed changes in stress-strain loops during cyclic loading are the result of slip band formation, GND accumulation, and the evolution of phase-field degradation. The resulting stored energy distributions and crack trajectories are given in Fig. \ref{fig:sx}b and \ref{fig:sx}d for the [100] and [111] studies, respectively.

First, let us consider the results obtained for the [100] orientation, Figs. \ref{fig:sx}a and \ref{fig:sx}b. Two clear regimes of behaviour are observed in terms of the crack propagation response. Initially, the distribution of plastic strain is symmetric with respect to the initial crack, and plasticity is mostly concentrated near the crack tip, resulting in a mode I crack growth trajectory, perpendicular to the loading direction. However, cyclic loading is leading to plastic accumulation and the formation of slip bands, as can be seen in the $G_s$ contour, Fig. \ref{fig:sx}a. As a result, the crack eventually deviates from the mode I trajectory to follow a crystallographic plane path along which it propagates due to the accumulation of stored energy density.

The results obtained when the load is applied along the [111] direction are shown in Figs. \ref{fig:sx}c and \ref{fig:sx}d. It can be seen that the distribution of plastic strain is asymmetric with respect to the crack plane, with the plastic strain and stored energy density $G_s$ concentrating on the upper-left and lower-right regions of the domain (see Fig. \ref{fig:sx}c). As shown in Fig. \ref{fig:sx}d, this results in mixed-mode cracking with the crack propagating along two crystallographic planes forming a 30$^{\circ}$ angle with the initial crack. These slip systems are thus found to be dominant relative to the other two slip systems in that plane, perpendicular and parallel to the initial crack.

Model predictions reveal crack trajectories that can be significantly influenced by the formation of slip bands along crystallographic planes. This is in agreement with experimental observations of crack path alignment with specific slip planes and deviation from the mode I trajectory at the micro-scale \cite{Rabbolini2014,Karamitos2022}.

\subsection{Failure of a bicrystal plate as a function of the grain boundary orientation}

\begin{figure}[ht!]
\centering
\includegraphics[width=0.7\textwidth]{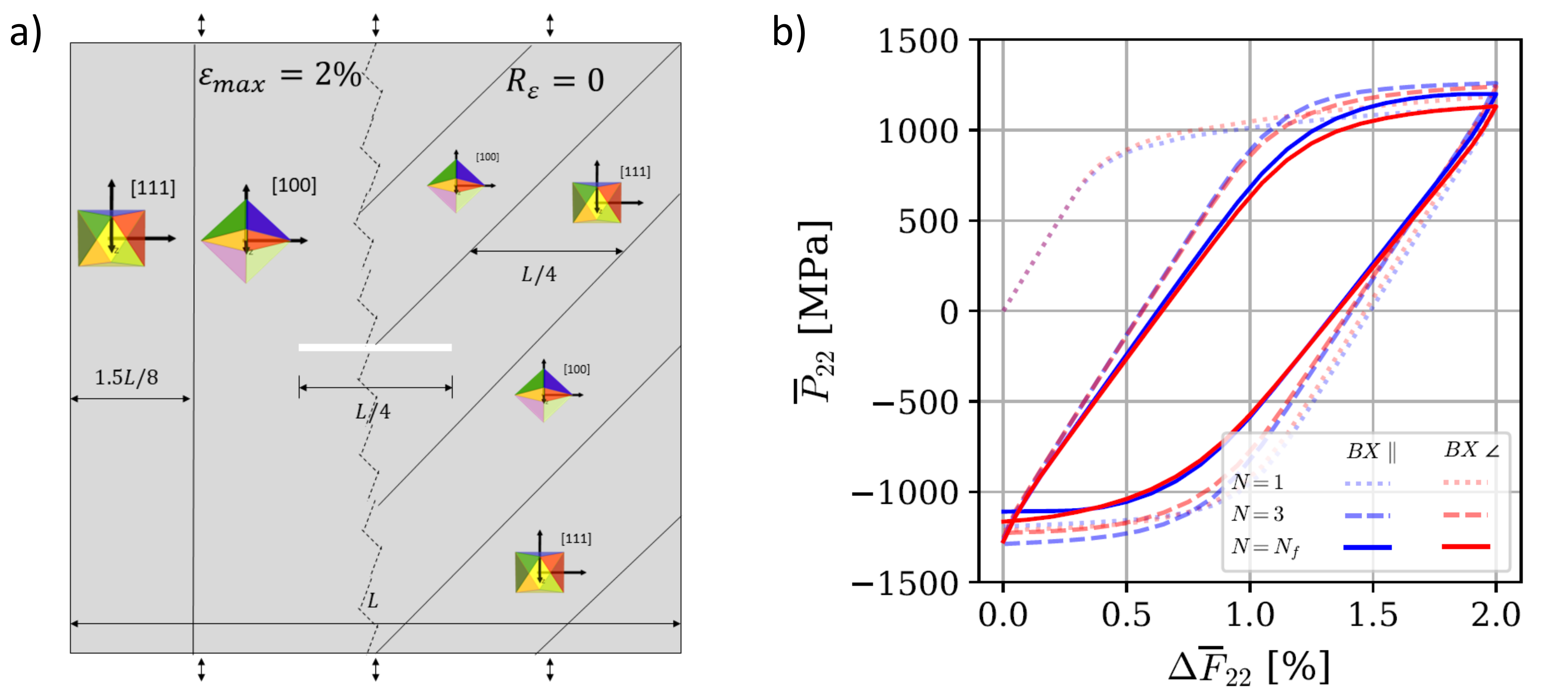}
\caption{Fatigue failure of a bicrystal plate as a function of the grain boundary orientation: (a) geometry of sample, boundary conditions and sketch of the crystal and grain boundary orientations, and (b) macroscopic stress-strain (2-vertical direction) results obtained for both case studies (parallel and inclined grain boundaries relative to the initial crack) at the first cycle, the third cycle and the failure cycle.}
\label{fig:bxgeo}
\end{figure}

The second case study aims at predicting the interplay between crack growth behaviour and grain boundary orientation. To this end, a bicrystal plate is investigated mimicking the geometry and boundary conditions of the previous study. That is, the plate dimensions are 200 $\upmu$m $\times$ 200 $\upmu$m $\times$ 0.78 $\upmu$, with the plate containing an initial crack of length 50 $\upmu$m. The same discretisation is also employed; 256 $\times$ 256 $\times$ 1 voxels. However, as depicted in Fig. \ref{fig:bxgeo}, this case study divides the domain into two crystallographic orientations, [100] and [111], and considers two case studies: (i) a sample with an inner [100] part and two outer [111] regions, with the grain boundaries aligned perpendicular to the initial crack, and (ii) a sample with alternating [100] and [111] regions and with the grain boundaries inclined 45$^{\circ}$ relative to the initial crack. It should be emphasised that in the present framework, grain boundaries constitute the interface between two domains with different properties, without any additional treatment or modelling assumption. This naturally leads to inhomogeneous plastic distributions and generation of GNDs near the grain boundaries.

\begin{figure}[ht!]\centering
\includegraphics[width=0.8\textwidth]{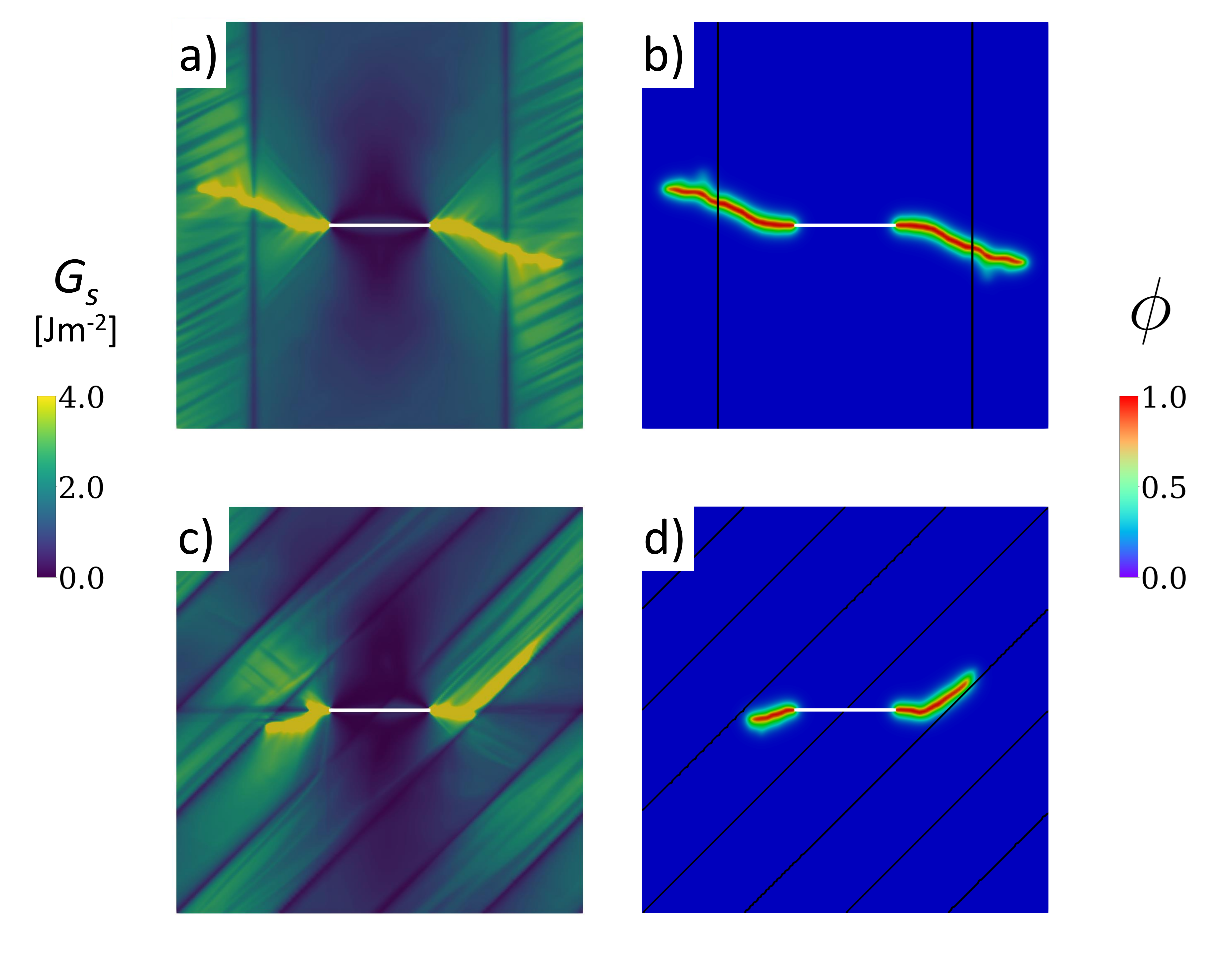}
\caption{Fatigue failure of a bicrystal plate as a function of the grain boundary orientation: perpendicular to the initial crack (top) versus inclined 45$^{\circ}$ relative to the initial crack (bottom). Subfigures (a) and (c) show the stored energy density $G_s$ contours, while subfigures (b) and (d) provide the predicted crack trajectories, as described by the phase-field contours. The results are shown at the end of the failure cycle $N_f$.}
\label{fig:bx}
\end{figure}

The macroscopic stress-strain responses obtained are shown in Fig. \ref{fig:bxgeo}b. As in the previous case study, the stabilised cyclic response is rapidly attained, requiring only three load cycles in both bicrystal analyses. Also, both scenarios reach the same number of cycles (5) before showing significant crack growth and a load drop. The grain boundary alignment appears to have a negligible influence on the macroscopic stress-strain response, as the two scenarios considered lead to almost identical stress-strain curves. However, a significant influence is observed in terms of microscopic crack trajectories, as shown in Fig. \ref{fig:bx}.

In both cases, as in the previous case study, cracking patterns deviate from the horizontal mode I trajectory that would be expected in isotropic materials. As the $G_s$ contours of Fig. \ref{fig:bx} reveal, this is due to plastic accumulation and the formation of slip bands. In addition, an interplay between dislocation densities and grain boundaries is also observed. Thus, the crack is able to propagate through the grain boundaries for the case study where the grain boundary is aligned perpendicular to the initial crack (Fig. \ref{fig:bx}a). However, when the grain boundary is inclined 45$^{\circ}$ relative to the initial defect, the crack trajectory is deflected and the crack grows along the grain boundary. As can be inferred from Fig. \ref{fig:bx}d, this is due to the interplay between dislocation density and grain morphology; plasticity is shown to be concentrated at some bands parallel to the grain boundaries in the [100] grains. Hence, grain boundaries are shown to act as barriers or preferential directions of crack growth. The results reveal the ability of the model in capturing the interplay between grain morphology, dislocation density and crack trajectory. 

\subsection{Nucleation and growth of fatigue cracks in a 2D polycrystal}

\begin{figure}[ht!]\centering
\includegraphics[width=0.7\textwidth]{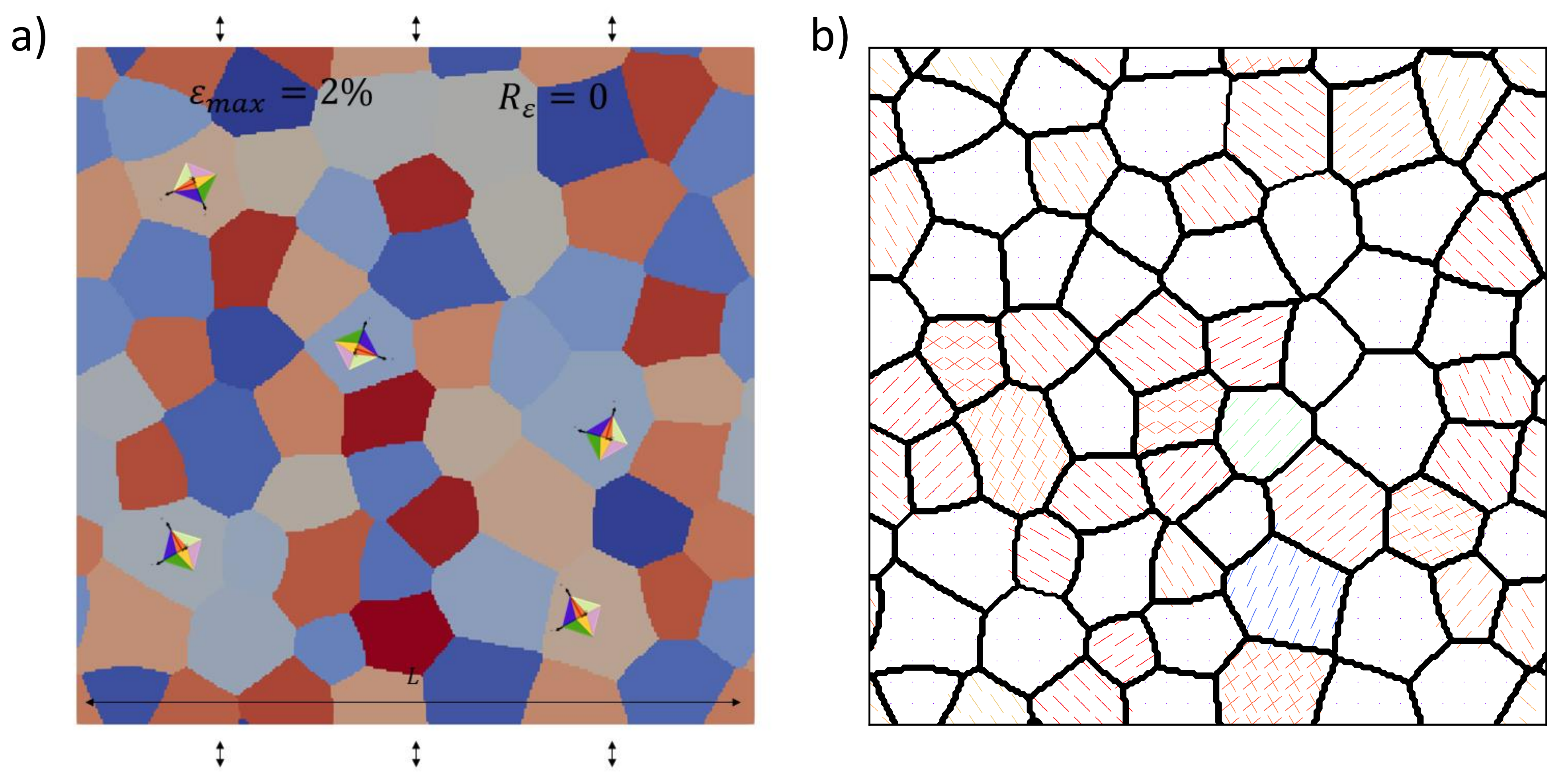}
\caption{Nucleation and growth of cracks in a polycrystal: (a) geometry, boundary conditions and grain distribution, and (b) sample microstructure, highlighting the slip plane projection lines (right) with Schmid factor $>$ 0.35.}
\label{fig:px1}
\end{figure}

\begin{figure}[ht!]\centering
\includegraphics[width=0.9\textwidth]{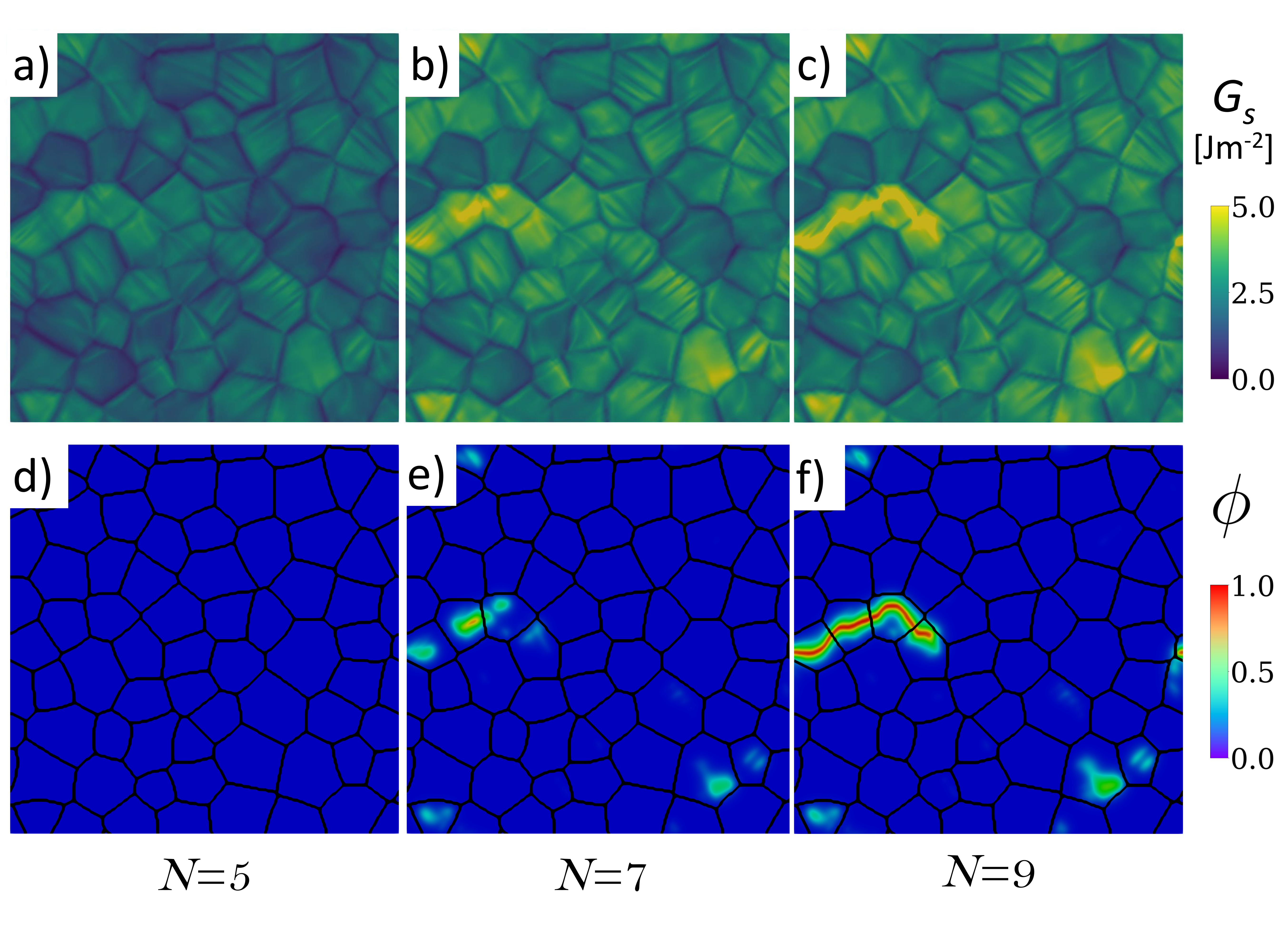}
\caption{Nucleation and growth of cracks in a polycrystal: predictions of stored energy density $G_s$ (top) and crack trajectories, as described by the phase-field contours (bottom). The results are shown as a function of the number of cycles $N$.}
\label{fig:px1h}
\end{figure}

The third case study aims at shedding light on both nucleation and short crack propagation of microstructural fatigue cracks. A polycrystalline plate of dimensions 200 $\upmu$m $\times$ 200 $\upmu$m $\times$ 0.78 $\upmu$m is considered, assuming plane stress behaviour and undergoing the cyclic boundary conditions described before. No initial crack is introduced, so as to study the nucleation process. The computational domain is discretised using 256 $\times$ 256 $\times$ 1 voxels. The plate contains a total of 62 crystals, which are randomly oriented and follow a log-normal distribution with a mean equivalent diameter of 30 $\upmu$m and a 0.1 scatter in standard deviation. The periodic microstructure studied is shown in Fig. \ref{fig:px1}, where the slip plane projection lines with Schmid factor larger than 0.35 are also given. The results obtained are given in Fig. \ref{fig:px1h}, in terms of plastic accumulation ($G_s$ contours) and crack trajectories ($\phi$ contours), as a function of the number of cycles. 

In agreement with the hypotheses of the model, the nucleation of microstructural fatigue cracks is driven by the accumulation of stored energy density. As shown in Fig. \ref{fig:px1h}, no damage is observed in the first cycles but the stored energy is being accumulated in slip bands inclined with respect to the loading direction within the grains. In this regard, it should be noted that polycrystalline configurations show a significant GND effect near grain boundaries, leading to lower plastic deformation due to grain boundary GND hardening and therefore low $G_s$ values. Thus, the model produces transgranular cracking with cracks originated within the grains and not at the grain boundaries. In order to account for non-crystallographic cracking, a modified definition of critical stored energy density parameter based on the specific microstructural features or environmental conditions of the material under investigation may be necessary.

The crack incubation process is shown to take several cycles and the nucleation of the primary crack occurs within one crystallographic plane of one of the grains with a high Schmid factor. Additionally, other damaged zones can be observed in Fig. \ref{fig:px1h}e, where regions with high dislocation activity are observed, which act as potential secondary crack formation sites. The results showcase how the present framework, which builds upon a phase-field description of fracture, can appropriately simulate crack formation. 

Following the nucleation of fatigue cracks, the microstructurally short crack propagation process takes place in a few cycles, as shown in Fig. \ref{fig:px1h}f. The crack follows specific crystal facets in some cases, but also deflects within some grains as a result of the contributions of multiple slip systems, showing in a polycrystalline setting the behaviour observed in the previous case studies. For instance, two active slip systems are shown on the right end of the crack in Fig. \ref{fig:px1h}c. After the crack spans several grains, the propagation becomes unstable due to periodicity and effective area reduction.

\subsection{Nucleation and growth of fatigue cracks in a 3D polycrystal}

\begin{figure}[ht!]\centering
\includegraphics[width=0.9\textwidth]{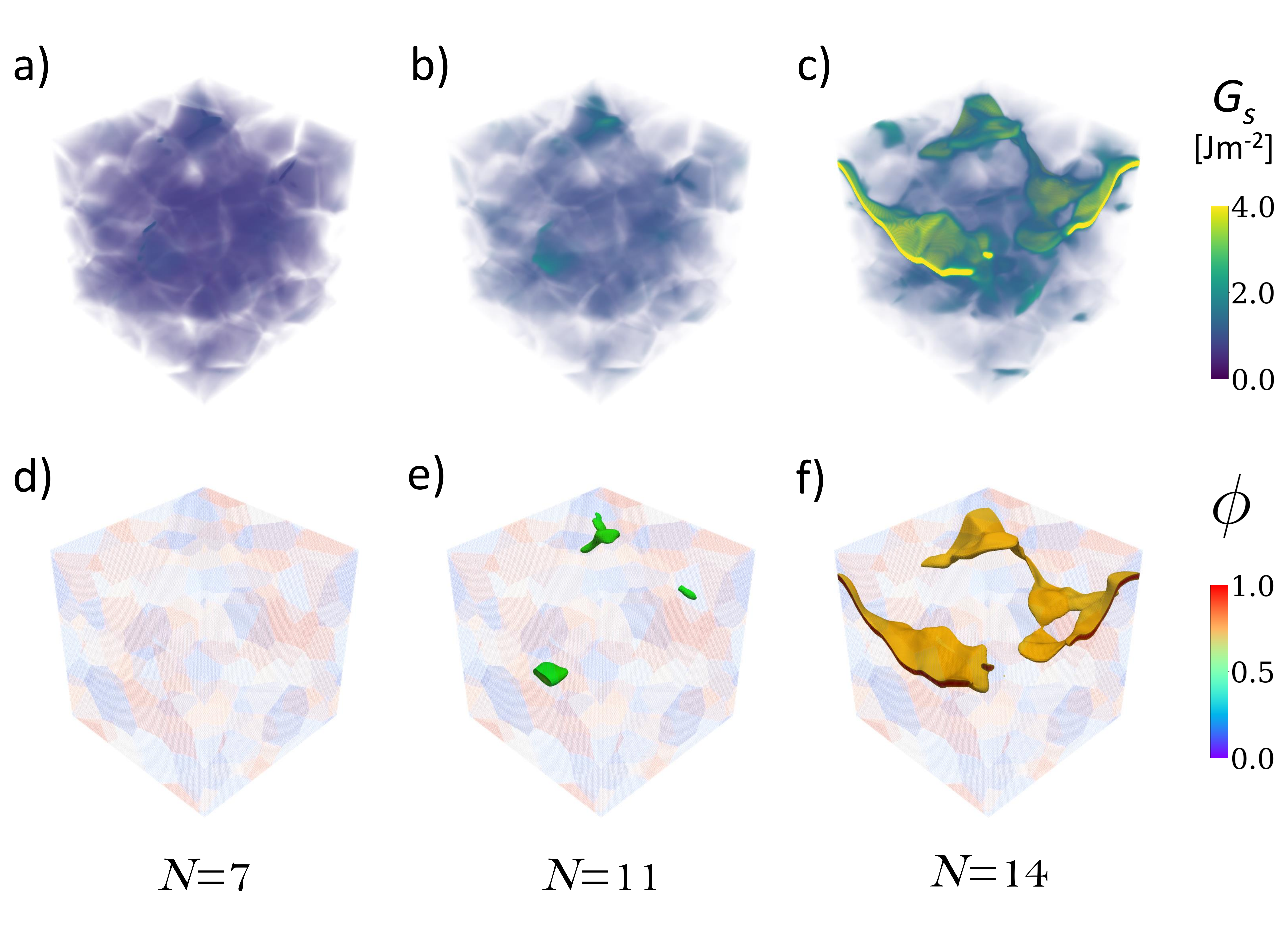}
\caption{Nucleation and growth of cracks in a 3D polycrystal: predictions of stored energy density $G_s$ (top) and crack trajectories, as described by the phase-field contours (bottom). The results are shown as a function of the number of cycles $N$. For ease of visualisation, phase field contours are shown only above a specific threshold value, equal 0.5 in subfigure (e) and of 0.8 in subfigure (f).}
\label{fig:p3dh}
\end{figure}

The fourth numerical experiment demonstrates the abilities of the modelling framework in predicting complex microstructural cracking phenomena in realistic 3D microstructures, taking advantage of the computation speed-up provided by FFT methods. Specifically, a cubic RVE containing 64 grains is considered. The RVE has dimensions of 100 $\upmu$m $\times$ 100 $\upmu$m $\times$ 100 $\upmu$m and is discretised using 128 $\times$ 128 $\times$ 128 voxels, which results in approximately $8\times10^6$ degrees-of-freedom. The synthetically generated periodic microstructure follows a log-normal distribution with an average diameter of 30 $\upmu$m and a standard deviation of 0.1. The RVE is subjected to the same cyclic loading conditions as the previous case studies, and the analysis is conducted until unstable cracking is observed. Model predictions are shown in Fig. \ref{fig:p3dh}, in terms of stored energy density $G_s$ and phase field $\phi$ contours as a function of the number of loading cycles $N$.

The results show that 11 cycles are required to accumulate sufficient levels of stored energy so as to trigger microstructural damage. Then, fatigue damage spreads relatively fast and multiple cracks are observed after 14 cycles. The model appropriately captures the interplay between plastic localisation and microstructural cracking. Prior to crack nucleation, the stored energy density parameter accumulates on slip planes within the grains and, similar to the 2D polycrystal case study, a significant effect of GNDs in preventing crack nucleation on grain boundaries is observed (see Fig. \ref{fig:p3dh}b). Nucleation sites can be detected from the first cycles (see Figs. \ref{fig:p3dh}a and \ref{fig:p3dh}b), as they are located where the highest levels of stored energy density parameter are attained. Damage is found to initiate in three locations (see Fig. \ref{fig:p3dh}e), resulting in the nucleation of three cracks that go on to propagate and interact. Again, as in the previous case study, crack nucleation sites are located within crystals with high Schmid factors. 

Cracking trends are shown to be closely connected with microstructural plastic localisation phenomena. Crack nucleation is found to follow crystallographic planes, and the subsequent propagation takes place throughout the microstructure following a transgranular cracking mechanism. The propagation trends observed are either single slip mode, following one plane, or multiple slip mode, a competition between two directions which was also observed in the previous 2D analyses. On the final cycles, the primary and secondary cracks coalesce due to their proximity (see Fig. \ref{fig:p3dh}f), leading to final crack patterns similar to those observed in experiments \citep{Rovinelli2018}. Eventually, multiple microstructural cracks interact and fracture becomes unstable. The results show the ability of the model in capturing the simultaneous nucleation of multiple cracks and complex cracking phenomena such as crack branching and the merging of multiple cracks. In addition, this case study demonstrates the ability of the framework to tackle large-scale 3D problems, which can be prohibitive with conventional finite element approaches. However, one must note that FFT solvers might suffer from scalability issues or high phase contrast limitations.

\subsection{Micro- to macro-structural crack growth in a notched sample with free surfaces}

\begin{figure}[ht!]\centering
\includegraphics[width=1\textwidth]{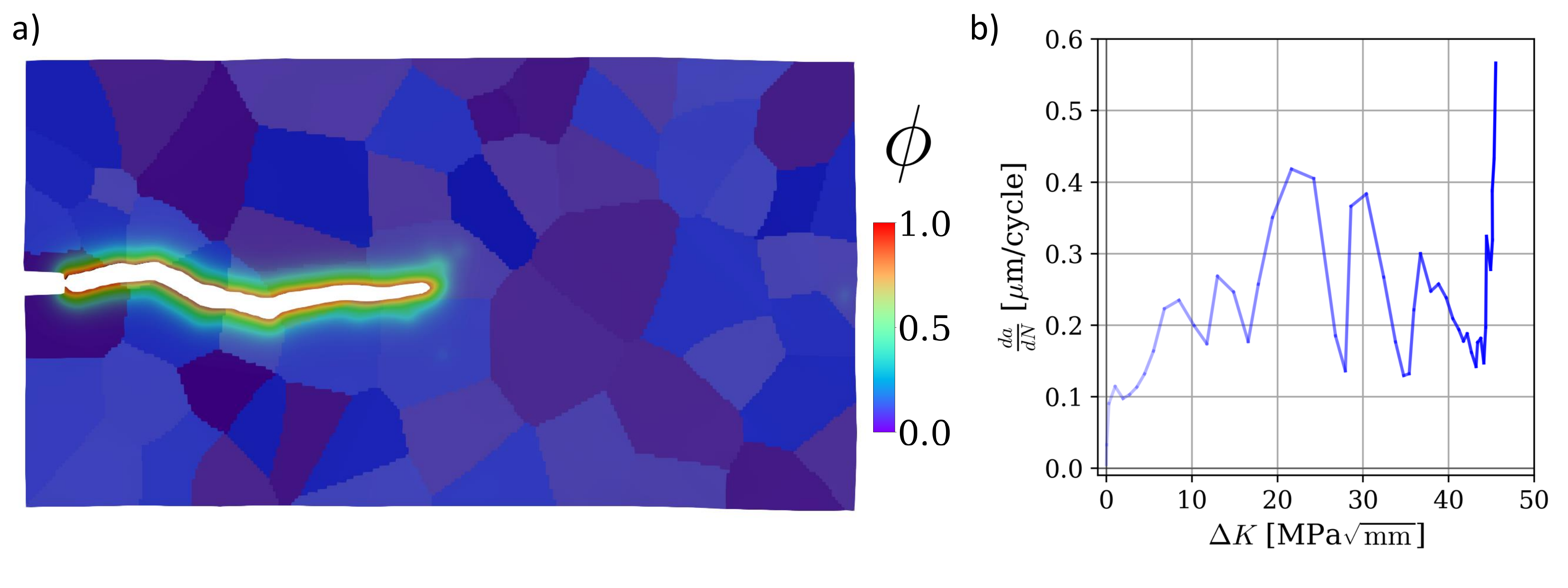}
\caption{Micro to macro crack growth in a notched sample: (a) crack propagation results ($\phi$ contours, removing regions of $\phi>0.9$ and with a scale deformation factor of 5), and (b) fatigue crack growth rates $da/dN$ as a function of the stress intensity factor amplitude $\Delta K$.}
\label{fig:not}
\end{figure}

Finally, we examine the propagation of a crack from a notch-like defect in a non-periodic boundary value problem where a dominant crack propagates along several grains. The aim is to observe a transition from micromechanical cracking behaviour, which is mostly dominated by plastic localisation and anisotropy, to macroscopic mode I crack growth, when the crack is sufficiently large relative to the relevant microstructural length scales. To this end, a sample of dimensions 200 $\upmu$m $\times$ 100 $\upmu$m $\times$ 0.78 $\upmu$m is considered, which is discretised using 256 $\times$ 128 $\times$ 1 voxels. As shown in Fig. \ref{fig:not}a, two free surfaces are introduced on both sides of the domain by including a buffer layer of voxels. In addition, a small notch is introduced on the left edge, effectively facilitating the nucleation of a crack and forcing the crack to propagate in only one direction. For both the free surface buffer layer and the notch region, a stiffness $10^{-5}$ times softer than the original crystal stiffness has been considered. It should be noted that the consideration of free surfaces is not suitable for standard FFT methods, here requiring an increase in the tolerance value to $tol_{lin}=10^{-3}$ to achieve convergence of the Krylov solver. The underlying microstructure and grain sizes are chosen so as to ensure that the width of the specimen spans 10 grains. This requires slightly reducing the grain size to keep the same voxel and phase field length scale magnitudes. The material parameters and loading conditions are kept the same as in previous case studies except for the assumptions of a load amplitude of $\varepsilon_{max}=1\%$ and a critical stored energy parameter of $G_{crit}=2$ Jm$^{-2}$. The results obtained are given in Fig. 8, in terms of the crack trajectory (removing regions where $\phi>0.9$) and the $da/dN$ vs $\Delta K$ behaviour computed using the formula for an edge crack in a semi-infinite body.

The calculations show that it only takes 4 cycles for a crack to nucleate at the tip of the pre-existing notch but that 45 cycles are then required for the crack to propagate over several grains. Initially, the crack trajectory is influenced by crystallographic planes but the sensitivity to the microstructure appears to diminish as the crack increases in size, becoming closer to the macroscopic mode I propagation path. A transition from mode II or mixed-mode to mode I fracture is typically observed in the early stages of fatigue damage \citep{Kunkler2008}. Crack propagation rates, shown in Fig. \ref{fig:not}b, exhibit initial oscillations resulting from the change in crack growth velocity that occurs when the crack crosses a grain boundary (relative to propagation within the bulk of a crystal), followed by a decrease due to the strain-controlled loading conditions. These oscillations due to transgranular cracking are in agreement with predictions from theoretical microplasticity-based fatigue models \citep{Chowdhury2016}.

\section{Concluding remarks}
\label{Sec:Conclusions}

This work presents a microstructure-based phase-field fatigue framework able to capture the nucleation and propagation of fatigue cracks in polycrystalline materials. The theoretical elements of the model combine dislocation-based crystal plasticity, a phase-field description of damage evolution, and a new fracture driving force based on the stored energy density concept. This driving force definition incorporates microstructural information and is shown to appropriately capture plastic localisation phenomena. The numerical implementation is carried out using FFT-based solvers, which enables simulating sufficiently large RVEs undergoing multiple loading cycles. Five case studies are investigated to gain insight into the abilities of the model in capturing nucleation and propagation of fatigue cracks in single crystals, bicrystals and 2D and 3D polycrystals. These numerical experiments have been carefully chosen so as to evaluate relevant microstructural plasticity-fracture interactions. The main findings are:
\begin{itemize}
    \item Model predictions show that short fatigue crack trajectories can be significantly influenced by the formation of slip bands along crystallographic planes. Cracks are found to grow along single crystallographic planes but also exhibit a mixed-mode behaviour resulting from the influence of multiple slip planes.
    \item Crack nucleation is observed to occur in preferentially oriented slip planes. Also, crack nucleation at grain boundaries is seen to be precluded by the consideration of GND effects - dislocation hardening that hinders plastic localisation at grain boundaries.
    \item Depending on the grain boundary orientation, cracks are observed to readily propagate through grain boundaries or to deflect and grow along them. The latter is more readily observed when grain boundaries are inclined 45$^{\circ}$ relative to the initial crack while the former is seen when cracks are perpendicular to grain boundaries.
    \item The analysis of polycrystals reveals the formation of cracks in grains of high Schmid factors and a subsequent transgranular crack propagation, following crystallographic plane paths.
    \item By combining efficient FFT methods and phase-field fracture, the model is able to predict complex cracking phenomena such as crack branching, as well as the nucleation of multiple cracks and their interaction, in relevant 3D RVEs and over multiple loading cycles.
    \item The simulation of an edge-cracked sample shows how the effect of the microstructure diminishes as the crack length becomes large relative to the relevant microstructural scales. Also, the model is able to capture the oscillations in fatigue crack growth rates that result from the interaction of cracks with grain boundaries.
\end{itemize}

\section*{Acknowledgements}

\noindent Sergio Lucarini acknowledges financial support from the Marie Skłodowska-Curie Individual European Fellowship under the European Union's Horizon 2020 Framework Programme for Research and Innovation through the project SIMCOFAT (Grant agreement ID: 101031287). Emilio Mart{\' i}nez-Pa{\~ n}eda acknowledges financial support from UKRI’s Future Leaders Fellowship programme [grant MR/V024124/1].

\small
\bibliography{Lucarini2023}

\end{document}